\documentclass{aa} 

\usepackage{xcolor}
\usepackage{graphicx}
\usepackage{txfonts}
\usepackage{natbib,twoopt}
\usepackage[breaklinks=true]{hyperref} %% to avoid \citeads line fills
\usepackage{comment}

\newcommand{\lya}{Ly$\alpha$}
\newcommand{\hi}{H~{\small I}}

\graphicspath{{./}{figures/}}

\begin{document}

\title{Spatially-resolved Spectroscopic Analysis of \lya\ Haloes}

\subtitle{Radial Evolution of the \lya\ Line Profile out to 60~kpc}

%\titlerunning{Radial Evolution of the \lya\ Line Profile}

\author{Yucheng Guo\inst{\ref{inst1}\thanks{e-mail: yucheng.guo@univ-lyon1.fr}}
\and Roland Bacon\inst{\ref{inst1}}
\and Lutz Wisotzki\inst{\ref{inst2}}
\and Thibault Garel\inst{\ref{inst3}}
\and J\'er\'emy Blaizot\inst{\ref{inst1}}
\and Joop Schaye\inst{\ref{inst4}}
\and Jorryt Matthee\inst{\ref{inst5}}
\and Floriane Leclercq\inst{\ref{inst11}}
\and Leindert Boogaard\inst{\ref{inst6}}
\and Johan Richard\inst{\ref{inst1}}
\and Anne Verhamme\inst{\ref{inst10}}
\and Jarle Brinchmann\inst{\ref{inst4}, \ref{inst8}, \ref{inst9}}
\and L\'eo Michel-Dansac\inst{\ref{inst1}}
\and Haruka Kusakabe\inst{\ref{inst7}}
}
%1
\institute{Univ Lyon, Univ Lyon1, Ens de Lyon, CNRS, Centre de Recherche Astrophysique de Lyon UMR5574, F-69230, Saint-Genis-Laval, France\label{inst1}
%2
\and Leibniz-Institut fur Astrophysik Potsdam (AIP), An der Sternwarte 16, 14482 Potsdam, Germany\label{inst2}
%3
\and Observatoire de Geneve, Universite de Geneve, 51 Ch. des Maillettes, 1290 Versoix, Switzerland\label{inst3}
%4
\and Leiden Observatory, Leiden University, P.O. Box 9513, 2300 RA Leiden, The Netherlands\label{inst4}
%5
\and Department of Physics, ETH Zürich, Wolfgang-Pauli-Strasse 27, 8093 Zürich, Switzerland\label{inst5}
%6
\and Department of Astronomy, University of Texas at Austin, 2515 Speedway, Austin, TX 78712, USA \label{inst11}
%7
\and Max Planck Institute for Astronomy, K\"onigstuhl 17, 69117, Heidelberg, Germany\label{inst6}
%8
\and Observatoire de Gen\`eve, Universit\'e de Gen\`eve, Ch. Pegasi 51, CH-1290 Versoix, Switzerland \label{inst10}
%9,10
\and Instituto de Astrofísica e Ciências do Espaço, Universidade do Porto, CAUP, Rua das Estrelas, PT4150-762 Porto, Portugal \label{inst8}
\and Departamento de Física e Astronomia, Faculdade de Ciências, Universidade do Porto, Rua do Campo Alegre 687, PT4169-007 Porto, Portugal \label{inst9}
%11
\and National Astronomical Observatory of Japan (NAOJ), 2-21-1, Osawa, Mitaka, Tokyo, 181-8588, Japan \label{inst7}
}

\date{Submitted 2023}
% \abstract{}{}{}{}{}
% 5 {} token are mandatory
\abstract
{
The extended \lya\ haloes (LAHs) have been found to be prevalent around high-redshift star-forming galaxies. However, the origin of the LAHs is still a subject of debate.
Spatially resolved analysis of \lya\ profiles provides an important diagnostic.
We analyse the average spatial extent and spectral variation of the circumgalactic LAHs by stacking a sample of 155 \lya\ emitters (LAEs) at redshift $3<z<4$ in the MUSE Extremely Deep Field.
Our analysis reveals that, with respect to the \lya\ line of the target LAE, the peak of the \lya\ line at large distances becomes increasingly more blueshifted up to a projected distance of 60~kpc ($\approx 3 \times$ virial radius), with a velocity offset of $\approx$ 250~km/s.
This trend is evident in both the mean and median stacks, suggesting that it is a general property of our LAE sample, which typically has a \lya\ luminosity $\mathrm{\approx 10^{41.1} erg\,s^{-1}}$.
However, due to the absence of systemic redshift data, it remains unclear whether the \lya\ line peak at large projected distances is less redshifted compared to the inner regions or truly blueshifted with respect to the systemic velocity.
We explore various scenarios to explain the large-scale kinematics of the \lya\ line.
}

\maketitle
%
%________________________________________________________________
\nolinenumbers

%%%%%%%%%%%%%%%%%%%%%%%%%%%%%%%%%%%%%%%%%%%%%%%%%%%%%%%%%%%%%%%%%%%%
\section{Introduction} \label{sec:intro}
%%%%%%%%%%%%%%%%%%%%%%%%%%%%%%%%%%%%%%%%%%%%%%%%%%%%%%%%%%%%%%%%%%%%
The exchange of mass, momentum, and energy between a galaxy and the intergalactic medium (IGM) plays a crucial role in regulating galaxy evolution.
The circumgalactic medium (CGM), which forms the transition region between the galaxy and the IGM, offers a detailed record of the relevant physical processes at play \citep[e.g.][]{tumlinson17}.
Throughout the epoch of galaxy evolution, the CGM is continuously enriched and depleted by galactic outflows, inflows, and mergers.
Investigating the different components of the CGM and understanding their connection to the galaxy's star formation activity and interstellar medium (ISM) are of critical importance for understanding galaxy evolution \citep[e.g.][]{hopkins18,thompson24}.

The hydrogen \lya\ line is a powerful tool for tracing the galaxies at high redshift and mapping their CGM and IGM.
Large samples of star-forming galaxies at $z \gtrsim 2$ featured by strong \lya\ emission, known as \lya\ emitters (LAEs), have been efficiently identified in deep narrow band (NB) imaging \citep[e.g.][]{gronwall07,ouchi08,shibuya12,kikuta23} and spectroscopic observations \citep[e.g.][]{rauch08,guo20b,richard21}.%jiang17,mentuch23
The spatially-integrated \lya\ emission line is almost always redshifted relative to the systemic velocity \citep[e.g.][]{yamada12,chonis13}, which is thought to be due to the scattering of \lya\ photons in a galactic outflow \citep[e.g.][]{ahn03,erb14,shibuya14}.
The \lya\ emission extended at the circumgalactic scale, known as \lya\ haloes (LAHs), is detected individually \citep[e.g.][]{rauch08,wisotzki16} and through stacking \citep[e.g.][]{hayashino04,momose14,wisotzki18}.
The integral field unit (IFU) facilities, such as the ESO-VLT instrument Multi Unit Spectroscopic Explorer \citep[MUSE,][]{bacon10} and Keck Cosmic Web Imager \citep[KCWI,][]{martin10,morrissey12} have enhanced the efficiency of detecting LAHs around galaxies at $2 \lesssim z \lesssim 6$.
We now know that the star-forming galaxies are ubiquitously surrounded by LAHs that are observed to be several to tens of times larger than the stellar continuum, indicating the presence of a significant amount of neutral hydrogen in the circumgalactic gas \citep[e.g.][]{steidel11,ouchi20,kusakabe22}. 
Recent deep observations and stacking analyses have detected extended \lya\ emission around LAEs out to hundreds of kpc, achieving a surface brightness level of $\mathrm{\approx 10^{-20}\,erg\,s^{-1} \, cm^{-2} \, arcsec^{-2} }$ \citep[e.g.][]{kikuchi22,maja22a,guo23}.

Despite that most studies of LAHs focus on mapping the surface brightness of the extended \lya\ emission, the spatially resolved spectral variation is still poorly constrained. 
Observing the spectral variation of the LAHs is of vital importance in understanding their physical nature, because it is related to physical properties of the CGM, such as geometry, gas column density, covering fraction, kinematics, and dust content \citep[e.g.][]{verhamme06,dijkstra14b,chang23}.
IFU observations of LAHs have so far been performed only on a few gravitationally lensed \citep{Patricio16,claeyssens19,solimano22} and unlensed \citep{erb18,erb22,leclercq20} LAHs, which, restricted by the S/N, can only study the luminous outliers at small impact parameters (10-30~kpc). 
In terms of how the \lya\ profile varies with distance, \citet{claeyssens19} and \citet{leclercq20} found a trend that the peak of the \lya\ line is shifted to redder wavelengths at large distances ($\approx$20~kpc) from the central galaxy, and that there is a strong correlation between this peak velocity shift and line width, which can be explained by models of \lya\ resonant scattering in outflows.
%A similar trend was also observed in the LAH of a strongly lensed galaxy pair by \citet{solimano22}. 
However, several other observations give contradictory results.
For example, in a single case study of an LAE with a particular double-peaked \lya\ profile, \citet{erb18} found higher blue-to-red peak ratios and narrower separations of the two peaks at larger radii ($\lesssim $30~kpc). \citet{erb22} extended the study by analysing a sample of 12 double-peaked LAEs, and found similar trends in peak ratios and separations as \citet{erb18}. 
Models have been engaged in reproducing the spatially resolved \lya\ profiles \citep[e.g.][]{zheng02,song20}.
From the simulation point of view, the variety of the \lya\ line shapes depends not only on the complicated radiative transfer effect in the galaxy's ISM and CGM, but also on the galaxy's evolutionary phase and the positional angle looking at the galaxy \citep{blaizot23}.

Despite previous spectroscopic observations on bright LAHs, the characteristic pattern in the spatial variations of the \lya\ line profile across the LAH population remains unclear.
To answer this question requires a large sample of LAEs to average out the variation between individual LAEs and deep observations to detect the very low surface brightness emission, which is difficult and expensive for current observing facilities.
We thus have to rely on data stacking.
The pioneering MUSE stacking analysis \citep{gallego21} suggests that the peak of the \lya\ emission is bluer at large radii, although with high noise in the spectra. 
More representative samples of LAEs are needed to investigate into the faint LAE population.
Higher S/N spectra are also needed to precisely understand the average \lya\ line profile and its velocity shift. 

In this work, we aim to obtain a comprehensive understanding of the average spectral variations of LAHs, particularly for the collective population of MUSE-detected LAEs, the majority of which have low \lya\ luminosity and a single redshifted \lya\ peak.
We make use of the dataset of MUSE eXtremely Deep Field survey \citep[MXDF,][]{bacon22}, which is the deepest spectroscopic datacube ever obtained. 
The MXDF is at least 3 to 10 times deeper than the datasets used in \citet{gallego21}, allowing for the identification of a population of much fainter LAEs and the detection of diffuse \lya\ emission at an unprecedentedly low surface brightness limit.
In our previous work \citep[][hereafter Paper~I]{guo23}, we measured the median \lya\ surface brightness profiles out to 270~kpc.
In this paper, we focus on the evolution of the \lya\ spectral profiles with distance.
We adopt the standard $\mathrm{\Lambda CDM}$ cosmology with $H_0\mathrm{=70\, km\, s^{-1}\, Mpc^{-1}}$, $\mathrm{\Omega _m=0.3}$ and $\mathrm{\Omega _\lambda =0.7}$. All distances are proper, unless noted otherwise.

%%%%%%%%%%%%%%%%%%%%%%%%%%%%%%%%%%%%%%%%%%%%%%%%%%%%%%%%%%%%%%%%%%%%
\section{Data description and analysis} \label{sec:data}
%%%%%%%%%%%%%%%%%%%%%%%%%%%%%%%%%%%%%%%%%%%%%%%%%%%%%%%%%%%%%%%%%%%%

The MXDF \citep{bacon22} has a field of view of 1' in diameter, with the longest exposure time of approximately 140 hours, and the shortest exposure time of several hours, reaching an unresolved emission line median $\mathrm{ 1\sigma }$ surface brightness limit of $\mathrm{ <10^{-19}\,erg\,s^{-1} \, cm^{-2} \, arcsec^{-2} } $.
The very deep exposure of the MXDF enables studies of extremely low surface brightness emission, such as the detection of a cosmic web in \lya\ emission on scales of several cMpc \citep{bacon21}.
The survey design, sky coverage and data reduction of the MXDF are described in \citet{bacon22}.
\citet{bacon22} also provide the redshifts, multi-band photometry, morphological and spectral properties, as well as measurements of the stellar mass and star formation rate of all the galaxies discovered by the deep MUSE observations in the Hubble Ultra Deep Field (HUDF).

A total of 420 LAEs have been detected in the MXDF at $3 \lesssim z \lesssim 6.5$.
In this study, we focus on LAEs in the redshift range $3<z<4$, where the cosmological surface brightness dimming is relatively weak, the IGM absorption is low, the efficiency of MUSE is high, and there are no strong atmospheric OH emissions.
There are 155 LAEs in this redshift range.
The distribution of redshift, \lya\ luminosity, \lya\ line rest-frame FWHM, and S/N of \lya\ line of the LAE sample can be found in Fig.~1 of Paper~I.
The median redshift is 3.41.
The median \lya\ luminosity ($\mathrm{L_{Ly\alpha}}$) of these LAEs is approximately $\mathrm{ 10^{41.1} erg\,s^{-1}}$, with the 90th percentile $\mathrm{ 10^{41.8} erg\,s^{-1}}$ and the 10th percentile $\mathrm{ 10^{40.6} erg\,s^{-1}}$.
The median stellar mass ($\mathrm{M_*}$) is approximately $\mathrm{ 10^{7.6} M_\odot}$. 
These characteristics make our sample representative of the fainter and lower mass populations of LAE compared to previous studies \citep[e.g.][]{leclercq20,gallego21,erb22}.
More details of this LAE sample are presented in \citet{bacon22} and Paper~I.
In Paper~I, we estimate the typical virial radius ($r_{vir}$) of this LAE sample based on the $r_{vir}$ - UV magnitude relation predicted by the semi-analytic model of \citet{garel15}.
The typical virial radius $r_{vir}$ of the sample is approximately 20~kpc.
The median $\mathrm{M_*}$ of the LAE sample also corresponds to a similar estimation of $r_{vir} \approx 23$~kpc, based on the stellar-to-halo mass relation given in \citet{girelli20}.
In the wavelength range corresponding to $3<z<4$, the spectral resolution of MUSE is approximately 150~km/s.

We produce a $21'' \times 21''$ MUSE mini-datacube centered on each LAE.
We mask all areas with an exposure time of less than 110 hours to achieve a high S/N.
We remove the continuum by performing spectral median filtering using a wide spectral window of 200~\AA.
This approach provides a fast and efficient way to remove continuum sources in the search for extended line emission, as presented in previous studies \citep[e.g.][]{wisotzki16,guo20,borisova16}.
We note that the major conclusions of this paper do not change if we change the spectral window of median filtering from 100~\AA\ to 300~\AA.
We also mask all the emission and absorption lines from neighbouring objects, based on the line catalogue and segmentation maps provided by \citet{bacon22}.

To study the average spatially-resolved spectra of the LAHs, we adopt a full 3D stacking procedure.
We shift individual mini-datacubes and re-bin them to a common (rest-frame) wavelength frame based on the peak wavelength of the \lya\ line. 
For LAEs with double-peaked \lya\ lines, we use the wavelength of the red peak.
We produce stacked datacubes using both mean and median stacking.

%%%%%%%%%%%%%%%%%%%%%%%%%%%%%%%%%%%%%%%%%%%%%%%%%%%%%%%%%%%%%%%%%%%%
\section{Results} \label{sec:SBprofile}
%%%%%%%%%%%%%%%%%%%%%%%%%%%%%%%%%%%%%%%%%%%%%%%%%%%%%%%%%%%%%%%%%%%%

%-------------------------------------------------------
\subsection{The large-scale kinematics of \lya\ line} \label{subsec:lya_shift}
%-------------------------------------------------------

%-------------------------------------------------------
%the lya nbs
%\begin{comment}
\begin{figure*}[t!]
\includegraphics[width=0.98\textwidth]{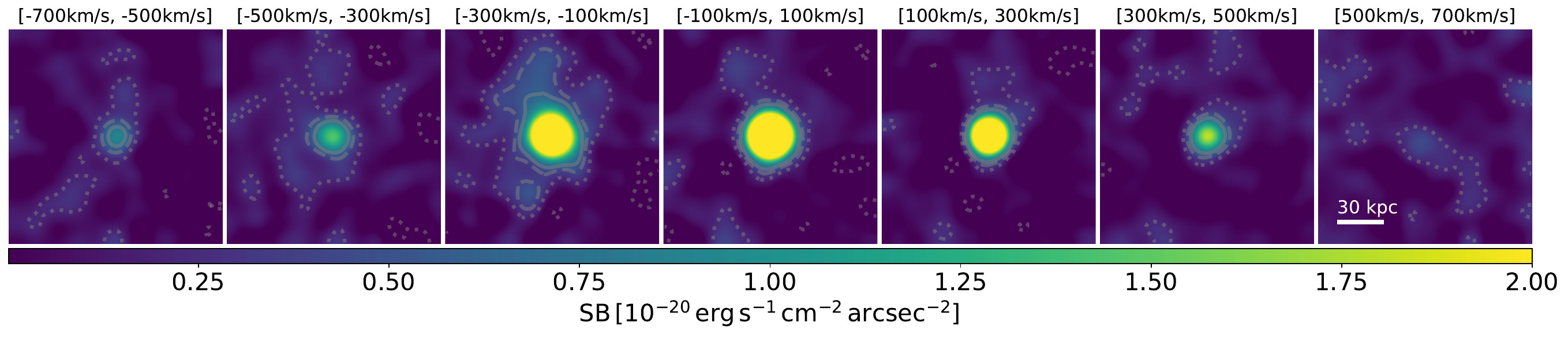}
\caption{Surface brightness maps of the median-stacked \lya\ emission.
Each panel shows the \lya\ pseudo-NB in a velocity interval of 200~km/s. The zero velocity corresponds to the peak of the \lya\ line. 
The NBs have been smoothed using a Gaussian kernel of width of $2.3''$.
The grey contours correspond to \lya\ significance levels of 2, 4 and 6 $\sigma$ (dotted, dashed and solid, respectively).
\label{fig_Lya_NBs}}
\end{figure*}
%\end{comment}

In Figure~\ref{fig_Lya_NBs} we present the \lya\ pseudo-NBs in a series of velocity bins of 200~km/s, with the full velocity coverage from -700~km/s to 700~km/s.
These \lya\ maps are extracted from the median-stacked datacube.
The width of the pseudo-NBs (200~km/s) corresponds to approximately 0.8~\AA\ at rest frame, and about 2-4 wavelength layers in the original MUSE datacubes. 
The middle velocity bin is centered on the \lya\ line peak of the central galaxy.
To improve the visualization of large-scale features, we spatially smoothed the NBs with a Gaussian filter with a FWHM of $2.3''$.

In Figure~\ref{fig_Lya_NBs} we clearly see that the \lya\ emission extends to several tens of kpc.
The most extended \lya\ emission does not reside at the peak of \lya\ line, but in the velocity range of [-300~km/s, -100~km/s], hinting that the extended \lya\ emission at large radii becomes bluer relative to the peak of the central \lya\ line.

%-------------------------------------------------------
%the lya line profiles
%\begin{comment}
\begin{figure}[ht!]
\includegraphics[width=0.48\textwidth]{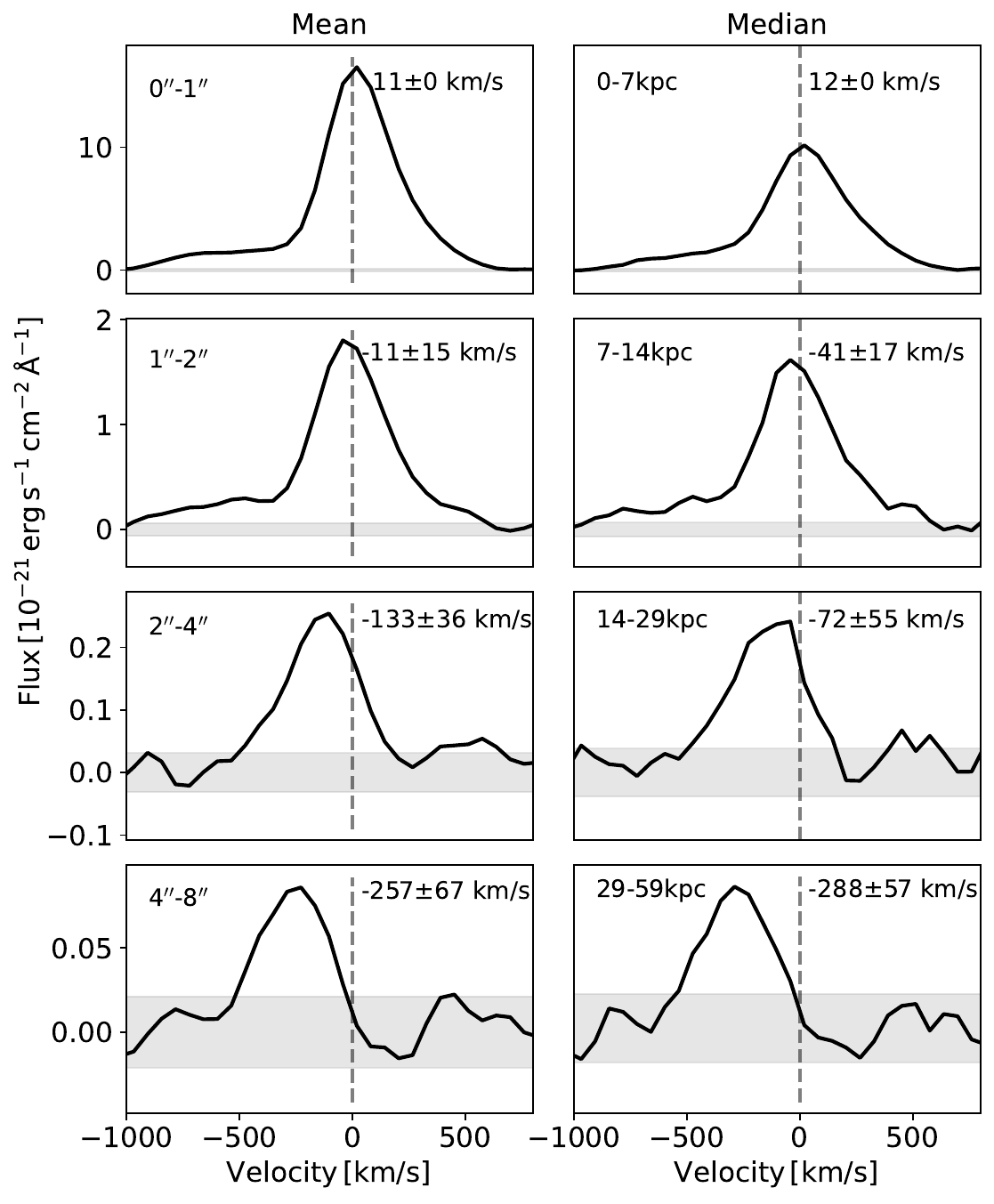}
\caption{\lya\ line profiles at different distances from the central galaxy. 
The spectra in the left and right columns are derived from the mean- and median-stacked datacubes. 
Different rows show different radial bins ranging from 1'' (7~kpc) to 8'' (59~kpc).
The vertical dashed lines denote the peak of the central \lya\ line. The horizontal grey shaded regions show the 1$\sigma$ error range.
\label{fig_Lya_profile}}
\end{figure}
%\end{comment}

%\begin{comment}
\begin{figure}[ht!]
\includegraphics[width=0.48\textwidth]{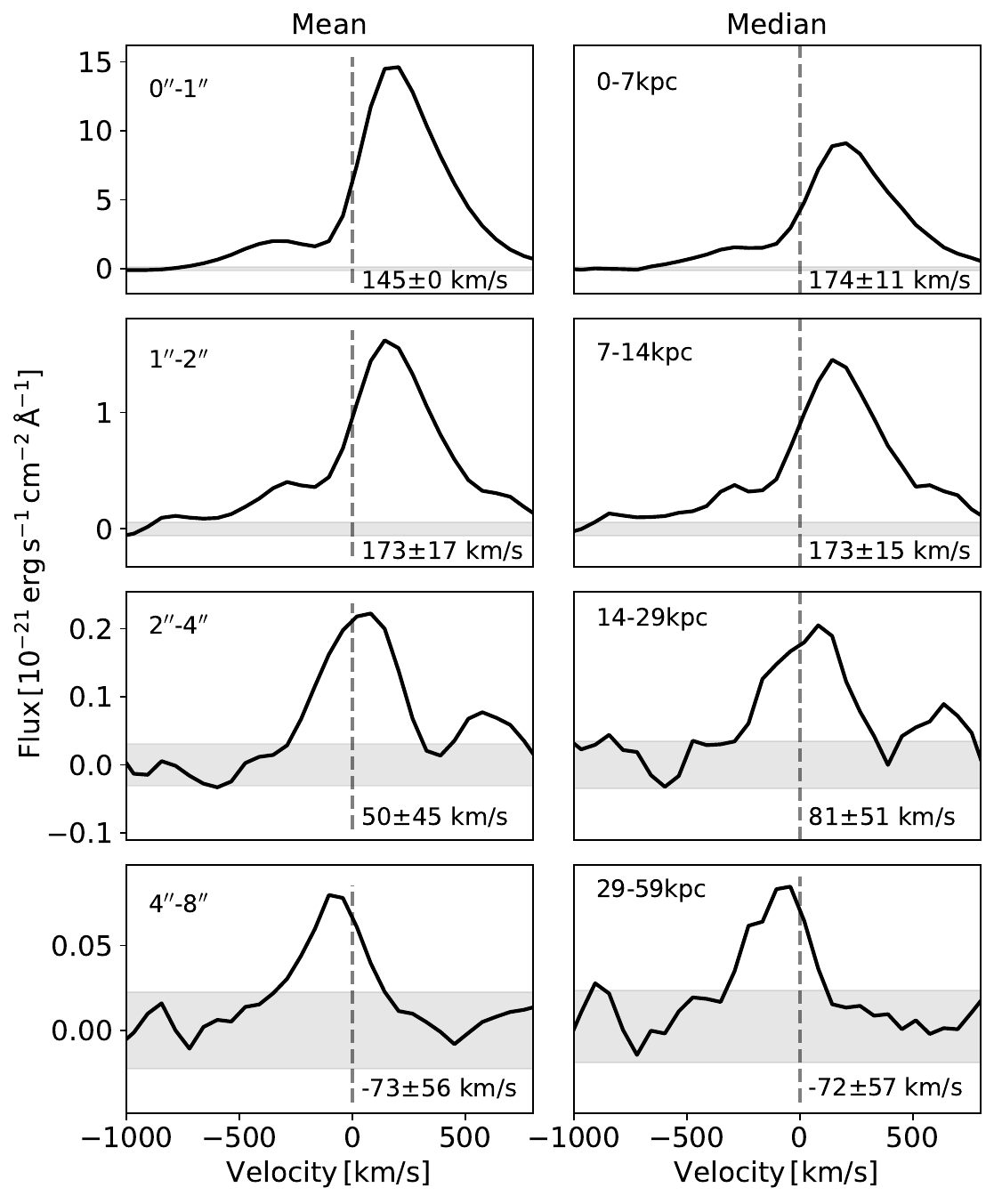}
\caption{
As Figure~\ref{fig_Lya_profile}, except that here all the mini-datacubes are re-aligned by the estimated $z_{sys}$ instead of $z_{Ly\alpha}$.
\label{fig_Lya_profile_zsys}}
\end{figure}
%\end{comment}

To confirm this blueshift trend of the extended \lya\ emission (with respect to the \lya\ red peak of the central galaxy), we measure the azimuthally averaged profiles of the \lya\ line.
The stacked spectra are shown in Figure~\ref{fig_Lya_profile}.
Different rows show different radial distances.
In both mean (left) and median (right), the \lya\ line shifts from approximately 0~km/s in the center to approximately -250~km/s at $\approx$60~kpc.
Note that the typical virial radius of the LAE sample is about 20~kpc, so we are observing the large-scale \lya\ kinematics out to approximately 3~$r_{vir}$.

This blueshift trend of the \lya\ line (with respect to the \lya\ peak of the central galaxy) naturally produces different behaviour of the \lya\ surface brightness profiles on the red and blue sides of the \lya\ line peak. As is presented in Figure~\ref{fig_Lya_sb_profile}, at large distances (tens of kpc), the bulk of the \lya\ emission is bluer than the \lya\ peak of the central galaxy, producing a flattening trend in the \lya\ surface brightness profile.
This flattening trend would be easily erased if the width of the pseudo-NB is too small.
In Paper~I, we measure the \lya\ surface brightness profile using reasonably wide pseudo NBs (with a width of 920~km/s at $z=3$), thus including most of the extended \lya\ emission.

%-------------------------------------------------------
\subsection{Error estimation} \label{subsec:lya_shift}
%-------------------------------------------------------

%-------------------------------------------------------
%noise
In Figure~\ref{fig_Lya_profile}, we derive the noise of the spectra using the velocity range of $\pm$1000-1500~km/s.
In the largest radial bin (29-59~kpc), the peak S/N of the \lya\ line is $\approx$3.2, and the accumulated S/N is $\approx$7.2.
For the spectrum in each radial bin, we reproduce 200 mock spectra based on the noise distribution.
We use these mock spectra to estimate the error of the \lya\ peak velocity. The values are shown in each panel of Figure~\ref{fig_Lya_profile}.

%-------------------------------------------------------
%error discussion
When we discuss the velocity shift, we must take into account the spectral resolution of MUSE, which is approximately 150~km/s in the wavelength range corresponding to $3<z<4$, smaller than the measured \lya\ line shift.
While the low spectral resolution may limit the measurement accuracy of the peak velocity, it cannot produce the smooth trend of \lya\ peak shift we see from 0~kpc to 60~kpc.
To further eliminate its influence, in Appendix~\ref{appendix:z45} we perform the stacking for LAEs at $4<z<5$ following the same procedures as in Section~\ref{sec:data}. In this wavelength range, the spectral resolution of MUSE is slightly higher \cite[approximately 110~km/s, see Fig.15 of][]{bacon17}. We see very similar trends as in Figure~\ref{fig_Lya_profile_z2}, but with lower S/N.

Another source of uncertainty comes from the measurement of the \lya\ peak wavelength of the central galaxy.
\citet{bacon22} provides the measurement errors for the \lya\ peak wavelength. The median value of this error over the LAE sample is approximately 165.1~km/s. 
When considering the average over 155 LAEs, the stacked error should be approximately 14~km/s, which is evidently smaller than the velocity shift observed for the largest radial bin. 

We also ensure that the stacking is not dominated by a few outliers. We perform the following analysis.
We randomly remove 15\%\ of objects in the sample, and repeat the data stacking procedure. 
This process is iterated 100 times, and for each stack, we measure the \lya\ peak velocity at 29-59 kpc.
The resulting median velocity and its standard error are approximately -259~km/s and 18~km/s, respectively.

%-------------------------------------------------------

%guidance for future obs
In summary, we detect a $\approx$250~km/s blueshift of the \lya\ line (compared to the \lya\ red peak of the central galaxy) from the galaxy out to 60~kpc ($3 \, r_{vir}$).
This blueshift trend is clearly observed in both mean and median stacks, and is not dominated by a few outliers.
Hence, it is a common phenomenon for the population of LAEs.
%Hence, it is a common phenomenon for LAEs with \lya\ luminosity $\mathrm{ 10^{41.1} erg\,s^{-1}}$.

%-------------------------------------------------------
\subsection{Where is the systemic velocity?} \label{subsec:z_sys}
%-------------------------------------------------------

In our previous analysis, we used the \lya\ peak redshift ($z_{Ly\alpha}$) of the target LAE as reference for spectral re-alignment and data stacking.
Due to the resonant nature of \lya\ photons, the \lya\ line usually deviates from the systemic redshift \citep[$z_{sys}$, e.g.][]{mclinden11,erb14,shibuya14,muzahid20}.
It is crucial to understand how the \lya\ line profile evolves radially with respect to the systemic velocity. However, in our sample, we lack direct measurements of $z_{sys}$.

We use the estimated $z_{sys}$ provided by \citet{bacon22}. 
This estimate is based on the empirical relation established by \citet{verhamme18}, which uses the \lya\ peak separation (if two peaks of the \lya\ line are detected) or the FWHM of the \lya\ line (if only one peak is detected) to estimate the velocity offset between $z_{Ly\alpha}$ and $z_{sys}$. 
We then re-align the mini-datacubes to the systemic velocity, and repeat the mean and median data stacking.
The \lya\ spectra in different radial bins are shown in Figure~\ref{fig_Lya_profile_zsys}.
In the inner $\approx$10~kpc, \lya\ is redshifted by about 170~km/s relative to $z_{sys}$.
With increasing distance, the \lya\ line shifts to shorter wavelengths.
For the final radial bin (29-59~kpc), the peak of the \lya\ line is slightly bluer than $z_{sys}$, with a value of $\approx$-70~km/s.
This trend can be seen in both the mean and median stacks.

The scatter of the $z_{sys}$ estimate of \citet{verhamme18} is about 72~km/s.
For a stack of 155 LAEs, the final error attributed to the $z_{sys}$ estimate should be only a few km/s.
However, due to the noise level and the resolution of the spectra, the \lya\ peak velocity in the final radial bin has a large uncertainty.
We cannot confirm whether the \lya\ peak at large distances (29-59~kpc) is really more blueshifted than the systemic velocity.

%%%%%%%%%%%%%%%%%%%%%%%%%%%%%%%%%%%%%%%%%%%%%%%%%%%%%%%%%%%%%%%%%%%%
\section{Discussion} \label{discussion}
%%%%%%%%%%%%%%%%%%%%%%%%%%%%%%%%%%%%%%%%%%%%%%%%%%%%%%%%%%%%%%%%%%%%
%\subsection{Comparison with previous works} 
\subsection{Average spectral morphology of the LAHs} 
\label{previous_works}

\begin{figure}[t!]
\includegraphics[width=0.48\textwidth]{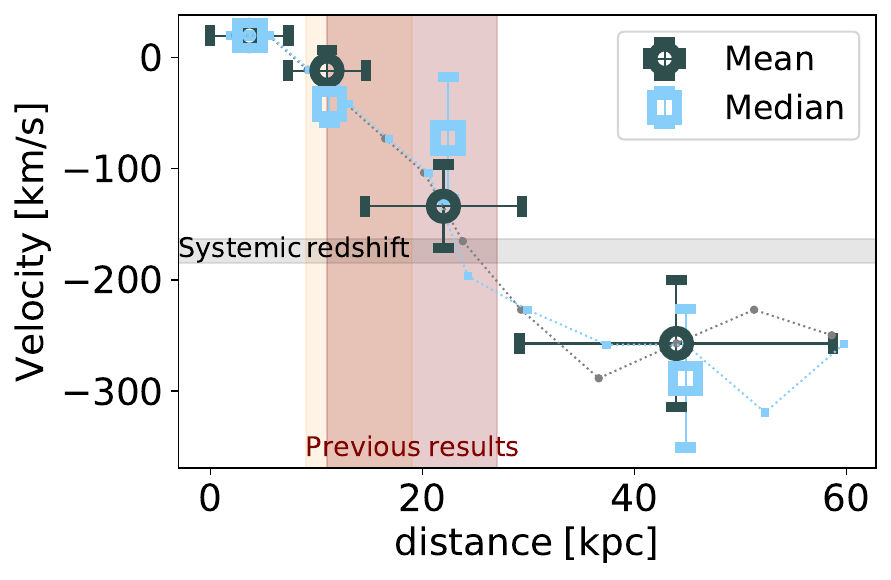}
\caption{
{The radial evolution of the \lya\ peak velocity with respect to the \lya\ line of the central LAE. 
The open circles and squares show the \lya\ peak velocity shift in the mean and median stacks (Figure~\ref{fig_Lya_profile}). 
The small symbols connected by dotted lines show the same results extracted with smaller radial bins.
The grey horizontal shadow shows the median systemic redshift of the central LAE, as is shown in Figure~\ref{fig_Lya_profile_zsys}, based on the empirical relations established by \citet{verhamme18}.
The orange and maroon vertical shadows show the \emph{maximum} radii of the LAHs from previous MUSE \citep{leclercq20,claeyssens19} and KCWI \citep{erb18,erb22} analysis.}
\label{fig_Lya_peak_vs_radius}}
\end{figure}

Characterising the spatially resolved spectroscopic features of the LAHs is important for understanding the physical nature of these systems.
Due to the faintness of the emission, previous work has mainly focused on a few bright or unique systems \citep[e.g.][]{claeyssens19,leclercq20,erb18,erb22}. 
They find that the spectral morphology of the extended \lya\ emission varies significantly between different LAHs, and even between different regions of an extended halo. 
However, a statistically robust general trend across the LAH population is still largely unknown.

In this work, we stack a sample of LAEs at $3<z<4$ observed by MUSE to study the typical spatial and spectral properties of the LAHs. We detect extended \lya\ emission out to the largest radius of $\approx$60~kpc (approximately 3~$r_{vir}$), and find an increasing blueshift trend of the \lya\ line with respect to the central \lya\ peak (Figures~\ref{fig_Lya_profile} and \ref{fig_Lya_profile_zsys}). 

In Figure~\ref{fig_Lya_peak_vs_radius}, we plot the \lya\ peak velocity (with respect to the central \lya\ peak) as a function of radial distance.
The open circles and squares with error bars represent the mean and median stacks in Figure~\ref{fig_Lya_profile}.
The small symbols connected by dotted lines show the same trend, but with smaller radial bins.
Although such a gradual trend of the \lya\ line shifting blueward with increasing radius is robust, we cannot confirm whether the \lya\ line at the largest radius (29-59~kpc) is truly bluer than the $z_{sys}$, due to the noise and the spectral resolution in the spectra, as is discussed in Section~\ref{subsec:z_sys}.

In Figure~\ref{fig_Lya_peak_vs_radius}, we also plot the maximum detection radii of the LAHs from the literature \citep{erb18,erb22,claeyssens19,leclercq20}. 
Before comparing our results with the literature works, we note that we detect \lya\ line profiles at much larger scales, not only in physical distance, but also with respect to the $r_{vir}$ of the system.
The LAEs in these literature works have much larger $r_{vir}$ than our LAE sample, because they are based on more luminous or more massive systems.
For example, the LAE sample of \citet{leclercq20} has a median UV magnitude of approximately -19.2~mag, which, based on the $r_{vir}$-UV magnitude relation given in \citet{leclercq17} and \citet{garel15}, corresponds to $r_{vir} \approx 50$~kpc.
The LAE sample of \citet{erb18,erb22} has a median stellar mass of $\mathrm{ 10^{9.13} M_\odot}$, corresponding to $r_{vir} \approx 70$~kpc, based on the stellar-to-halo mass relation given in \citet{girelli20}.
Although this estimate of $r_{vir}$ is subject to large uncertainties, it does provide an indication of the different physical scales we are discussing: 
The maximum detection radii of the LAHs in \citet{leclercq20} and \citet{erb22} are approximately 19~kpc and 27~kpc, respectively, which means that they observe the \lya\ lines within 50\% of the $r_{vir}$, namely, the inner CGM.
Our stack, however, reaches a much larger scale with respect to the typical virial radius of the system (60~kpc, or approximately 3~$r_{vir}$), thus we are analysing the physical processes that occur far beyond the inner CGM.

This large-scale velocity shift of the \lya\ line shown in Figure~\ref{fig_Lya_peak_vs_radius} does not agree well with several previous analyses of individual LAHs, such as \citet{claeyssens19} and \citet{leclercq20}.
They find a trend that the \lya\ line in the halo is redder compared to the \lya\ line in the centre, although the trend has a large scatter. 
Limited by the S/N, the maximum radii of the LAHs in \citet{claeyssens19} and \citet{leclercq20} are about 9-19~kpc.
As is shown in Figure~\ref{fig_Lya_peak_vs_radius}, the average \lya\ peak shift within this radius is not obvious. 
This weak signal is easily drowned out by the poor S/N and the intrinsic variation of the small sample.

The KCWI observations of \citet{erb18,erb22} on 13 double-peaked LAEs at $z\sim2.3$ find a similar trend to our work, that the \lya\ line becomes more blue dominated with increasing radii, up to a maximum distance of $\approx$18-26~kpc. 
An advantage of \citet{erb18,erb22} is that they have a secure measurement of $z_{sys}$ from rest-frame optical emission lines, so they actually find that the blue component (with respect to the systemic velocity) becomes more prominent at larger radii.
The LAEs in \citet{erb18,erb22} are specifically selected to have low metallicity and high ionisation by extreme nebular emission line ratios, or to have very strong \lya\ emission.
It is therefore unclear whether we can generalise their conclusions to the LAE population.
There are clear differences between our LAE sample and the LAEs in \citet{erb18,erb22}. 
Their LAEs have a 2~dex higher median $\mathrm{L_{Ly\alpha}}$ and a 1.5~dex higher median stellar mass than our sample.
%Their LAEs have a median $\mathrm{L_{Ly\alpha}}$ of $\mathrm{ \approx 10^{43.1} erg\,s^{-1}}$, 2~dex higher than the median of our sample, and a median stellar mass of $\mathrm{ 10^{9.13} M_\odot}$, 1.5~dex higher than the median of our sample.
These differences in turn imply that at they are probing the inner CGM ($\lesssim$0.5~$r_{vir}$), while we are probing the very extended CGM out to approximately 3~$r_{vir}$. This is likely the main origin of the difference in our results. 

All the LAEs in \citet{erb18,erb22} have double peaked \lya\ lines. 
Within the distance range 11-26~kpc, the LAHs in \citet{erb18,erb22} always have two individual \lya\ peaks on either side of the systemic velocity, while the relative strength of the two peaks changes with distance.
Similarly, at the distance $\lesssim$0.5~$r_{vir}$, our stack finds a broad \lya\ component bluer than the systemic velocity (top two rows of Figure~\ref{fig_Lya_profile_zsys}), which results from the stacking of many blue \lya\ lines at different velocities.
However, at the distance larger than 14~kpc, we observe only a single \lya\ peak.
This single peak cannot solely be attributed to the lower spectral resolution of MUSE:
The peak separation of \citet{erb18,erb22} at their largest radial bin ranges from approximately 300 to 800~km/s.
If the MXDF LAEs have a similar peak separation to \citet{erb18,erb22}, the two peaks could be resolved by MUSE.

The IGM absorption could remove the flux bluer than the systemic velocity, leading to an underestimation of the \lya\ blue peak flux.
This systematic suppression of the \lya\ blue peak with increasing redshift has been well analysed in previous work \citep[e.g.][]{hayes21}. 
At the median redshift $z\sim3.4$ of our LAE sample, the typical IGM transmission of \lya\ is
$\approx$~60\% \citep{laursen11,inoue14}.
The poorly constrained $z_{sys}$ complicates our analysis:  
At large distances (29-59~kpc in Figure~\ref{fig_Lya_profile_zsys}), if the \lya\ line is still redder than the systemic velocity, the \lya\ flux would not be strongly affected by the IGM transfer.
However, if it's the other way round, the part of the \lya\ flux that is bluer than the systemic velocity would be suppressed by IGM absorption by an average factor of 0.6.

\subsection{Physical nature of the LAHs} \label{mechnisms}

Several physical mechanisms have been proposed to explain the production and propagation of \lya\ photons in LAHs. 
Their relative importance may vary with radial distance \citep[e.g.][]{mitchell20,byrohl21}.
Here we explore several speculative considerations regarding the underlying physics responsible for the observed spectral shift of the \lya\ line out to a distance of 60~kpc (approximately 3~$r_{vir}$).
Notably, a successful physical model should also account for the observed \lya\ surface brightness profile, which shows a power-law decrease within 20~kpc and a tentative flat trend at 30 -- 50~kpc (paper~I).

%Outflow
The strongest peak of the \lya\ line in most LAEs is observed to be redshifted relative to the systemic velocity \citep[e.g.][]{yamada12,erb14},
which is commonly attributed to the complicated radiative transfer in outflows \citep[e.g.][]{verhamme06,dijkstra12,blaizot23}.
%which is commonly attributed to the complicated radiative transfer in outflows, for example, backscattering \citep[e.g.][]{verhamme06,dijkstra12} or highest optical depth in the velocity range bluer than the systemic velocity \citep[e.g.][]{blaizot23}.
Models of \lya\ photons scattering from a central source into an outflowing medium have successfully reproduced a large number of spatially integrated \lya\ spectra \citep[e.g.][]{hashimoto15,gronke17,chang23}.
Several previous works have been engaged in explaining the spatially resolved features of LAHs \citep[e.g.][]{zheng02,song20}.
To account for the observed large-scale blueshift trend of the \lya\ line with respect to the central LAE, one possibility is a gradually decelerating outflow,
which is expected because of the increasing importance of gravitational deceleration at large radii. 
The decelerating outflow could produce a \lya\ line with less redshifted velocity offsets with respect to systemic velocity at larger distances \citep[e.g.][Garel et al. in preparation]{song20}. 
However, the outflowing gas could hardly produce a \lya\ line bluer than systemic velocity.
Similarly, radiative transfer modelling of \citet{erb22} finds a gradual deceleration or constant-velocity phase of the outflow at large radii, which can recover the increase in \lya\ blue peak flux with distance. The LAEs of \citet{erb22} are double-peaked; most of their \lya\ blue peaks at large radii are fainter than the red ones.

The simulations of \citet{blaizot23} suggest an alternative scenario where the strongest resonant scattering effects (broadening and redshifting of the line) occur in the inner CGM where gas densities are high. Instead, the more extended CGM (between 0.2 and 3~$r_{vir}$) is often dominated in volume by a hot and highly ionised outflow, which acts as a screen that intercepts blue \lya\ photons from the galaxy and re-emits them more or less isotropically. 
This process is expected to produce a profile increasingly moving blueward at large distances with respect to the central \lya\ line. 

%-------------------------------------------------------
%inflows
Another scenario considers the role of inflows. \citet{chung19} simulate an LAH around a galaxy with a stellar mass of $\mathrm{10^{10.5} M_\odot}$ \citep[$r_{vir} \approx 95$~kpc based on][]{girelli20} and suggest that galactic outflows primarily affect the \lya\ properties within $\approx$50~kpc, while cold accretion flows dominate at larger radii. 
Based on the observation of an enormous \lya\ blob, \citet{li22} provide a model in which the \lya\ photons are produced by star formation in the galaxy and propagate outwards within multiphase clumpy outflows. 
Near the blob outskirts, infalling cool gas shapes the observed blue-dominated \lya\ line, albeit with a very small contribution to the total \lya\ luminosity.
This combination of outflows and inflows is also inferred from the \hi\ absorption by \citet{chen20}.
They observe an asymmetry in the \lya\ absorption line and explain it as infilling by blueshifted \lya\ emission relative to $z_{sys}$. 
They also claim a transition between outflow-dominated ($\mathrm{r \lesssim 50~kpc}$) and accretion-dominated flows ($\mathrm{r \gtrsim 100~kpc}$) for galaxies with $z \approx 2.2$ and $\mathrm{M_* \approx 10^{10} M_\odot}$ by comparing their observations with models.
If we assume that the \lya\ line at large distances (Figure~\ref{fig_Lya_profile_zsys}) is indeed blueshifted with respect to the systemic velocity, then a scenario of large-scale gas inflows becomes reasonable, in which a non-negligible fraction of \lya\ photons is propagated in gas inflows, and the gradual transition of the \lya\ line from redshift to blueshift out to tens of kpc indicates a transition in the dominance of gas outflows to inflows.

%-------------------------------------------------------
%\lya\ flux budget
From the perspective of the \lya\ flux budget, scattering models in which all \lya\ photons are scattered from a central source inherently predict a monotonically decreasing \lya\ surface brightness profile, while our observations (Appendix~\ref{appendix:sb_profile} and Paper~I) find a flattening trend at large distances. 
Paper~I suggest that this trend requires additional sources of \lya\ photons instead of scattering from the central galaxy, e.g., \textit{in situ} production of \lya\ photons at large distances.
The power source could be cooling radiation that converts gravitational energy into kinetic and thermal energy through collision \citep[e.g.][]{haiman00,dijkstra06a,rosdahl12}.%furlanetto05,dijkstra06b
Fluorescence from the nearby galaxies or from the UV background can also produce \lya\ photons at large radii \citep[e.g.][]{gould96,cantalupo05,ribas16}.
%furlanetto05  gallego18,gallego21
If the \textit{in situ}-produced \lya\ photons are emitted from a static medium and escape directly without scattering, the \lya\ line should be exactly at the systemic velocity.
However, if the bulk of the locally produced \lya\ photons are scattered in the CGM, the kinematics of the CGM gas would strongly influence the \lya\ line profiles, with the gas inflows and outflows producing blue- and red-skewed \lya\ profiles with respect to the systemic velocity.

%-------------------------------------------------------
%satellites
Neighbouring galaxies could also contribute to the observed properties of LAHs, whether they are unresolved satellite galaxies \citep[e.g.][]{momose16,mas17,mitchell20} or more massive nearby systems \citep[e.g.][]{byrohl21}.
Considering the \lya\ emission from the neighbouring galaxies is also regulated by similar mechanism(s) as the target galaxy, the \lya\ emission escaping from these galaxies should also averagely be redshifted relative to the systemic velocity.
Assuming either radial infalling or virialised motions of the satellite galaxies, it is difficult to find a physical model that would systematically produce a radial blueshift (or less-redshift) trend in the stack.
%Therefore, the neighbouring galaxies could hardly reproduce the observed \lya\ peak shift at large distances.
However, under the assumption that the unresolved satellites have lower star-formation rates and possibly drive weaker winds, we could expect smaller \lya\ velocity offsets from satellite galaxies \citep{muzahid20}.

%-------------------------------------------------------
\hspace*{\fill}

Overall, our observed \lya\ peak shift appears to be a generic property of LAHs according to the MXDF data.
A possible scenario would be that the \lya\ line emerging at small projected radii results from the propagation of centrally emitted photons in outflowing gas, whereas the signal at the outskirts is more likely due to radiative transfer effects in infalling gas or decelerating outflows.
Distinguishing between the two kinematic patterns requires a more robust measurement of $z_{sys}$, which is not available at this work.
The \lya\ photons at the outskirts are likely to be produced \textit{in situ} by collision, fluorescent emission or satellites.
To investigate the physical nature of the LAHs, a better understanding of both the flux budget and the kinematic behaviour of the LAHs is required.

%%%%%%%%%%%%%%%%%%%%%%%%%%%%%%%%%%%%%%%%%%%%%%%%%%%%%%%%%%%%%%%%%%%%
\section{Summary} \label{sec:summary}
%%%%%%%%%%%%%%%%%%%%%%%%%%%%%%%%%%%%%%%%%%%%%%%%%%%%%%%%%%%%%%%%%%%%

The main focus of this paper is to report a large-scale blueshift of diffuse \lya\ emission (with respect to the \lya\ line of the central LAE). 
This paper is based on a statistical analysis of a representative galaxy sample with a typical \lya\ luminosity of $\mathrm{ 10^{41.1} erg\,s^{-1}}$.
Our results represent a common phenomenon across the LAE population down to its faint end.
Thanks to the extremely deep MXDF survey and the data stacking, we are able to detect the \lya\ line profile out to 60~kpc (approximately 3~$r_{vir}$), much larger than any previous analyses based on a few bright LAEs \citep[e.g.][]{leclercq20,erb22}.
The main results are summarised below:

From the galaxy out to $\approx$ 60~kpc, we detect a $\approx$250~km/s blueshift of the \lya\ emission line with respect to the \lya\ peak of the target LAE.
However, we cannot determine the relative velocity offset between the \lya\ line at the large distance and the systemic velocity, so we cannot tell whether it is bluer or redder than $z_{sys}$.
We have demonstrated that the stacking of the LAE sample is not dominated by outliers. The blueshift trend with respect to the central \lya\ red peak is thus ubiquitous among LAEs at $3<z<4$ with a median \lya\ luminosity $\mathrm{L_{Ly\alpha} = 10^{41.1} erg\,s^{-1}}$.

We discuss the possible mechanisms to explain the blueshift of the \lya\ emission. Our observations favour a scenario in which the inner part of the LAH is dominated by resonant-scattered \lya\ photons from the outflowing gas, while at large distances, the outflows become gradually decelerated or inflows start to emerge out to $\approx$60~kpc.
However, at large distances, we cannot rule out the contribution from other scenarios, such as fluorescent emission or satellites.

Obviously, this work is limited by the absence of $z_{sys}$. The key to tackling this problem is to obtain the rest-frame optical spectra. Spatially and spectrally resolved analytical models and simulations are also needed for better interpretation of the observational data.

%%%%%%%%%%%%%%%%%%%%%%%%%%%%%%%%%%%%%%%%%%%%%%%%%%%%%%%%%%%%%%%%%%%%
\begin{acknowledgements}
%Y.G., R.B. acknowledge support from the ANR L-INTENSE (ANR-20-CE92-0015). 
Y.G., R.B. and L.W. acknowledge support from the ANR/DFG grant L-INTENSE (ANR-20-CE92-0015, DFG Wi 1369/31-1).
L.W. acknowledges support by the ERC Advanced Grant SPECMAG-CGM (GA101020943).
J.Brinchmann acknowledges support by Fundação para a Ciência e a Tecnologia (FCT) through the research grants UID/FIS/04434/2019, UIDB/04434/2020, UIDP/04434/2020 and PTDC/FIS-AST/4862/2020. 
J.Brinchmann acknowledges support from FCT work contract 2020.03379.CEECIND.
\end{acknowledgements}

\bibliographystyle{aa} % style aa.bst
\bibliography{ms} % your references Yourfile.bib

\begin{thebibliography}{71}
\expandafter\ifx\csname natexlab\endcsname\relax\def\natexlab#1{#1}\fi

\bibitem[{{Ahn} {et~al.}(2003){Ahn}, {Lee}, \& {Lee}}]{ahn03}
{Ahn}, S.-H., {Lee}, H.-W., \& {Lee}, H.~M. 2003, \mnras, 340, 863

\bibitem[{{Bacon} {et~al.}(2010){Bacon}, {Accardo}, {Adjali}, {Anwand}, \& {Bauer}}]{bacon10}
{Bacon}, R., {Accardo}, M., {Adjali}, L., {Anwand}, H., \& {Bauer}, S. 2010, in Society of Photo-Optical Instrumentation Engineers (SPIE) Conference Series, Vol. 7735, Ground-based and Airborne Instrumentation for Astronomy III, ed. I.~S. {McLean}, S.~K. {Ramsay}, \& H.~{Takami}, 773508

\bibitem[{{Bacon} {et~al.}(2023){Bacon}, {Brinchmann}, {Conseil}, {Maseda}, {Nanayakkara}, {Wendt}, {Bacher}, {Mary}, {Weilbacher}, {Krajnovi{\'c}}, {Boogaard}, {Bouch{\'e}}, {Contini}, {Epinat}, {Feltre}, {Guo}, {Herenz}, {Kollatschny}, {Kusakabe}, {Leclercq}, {Michel-Dansac}, {Pello}, {Richard}, {Roth}, {Salvignol}, {Schaye}, {Steinmetz}, {Tresse}, {Urrutia}, {Verhamme}, {Vitte}, {Wisotzki}, \& {Zoutendijk}}]{bacon22}
{Bacon}, R., {Brinchmann}, J., {Conseil}, S., {et~al.} 2023, \aap, 670, A4

\bibitem[{{Bacon} {et~al.}(2017){Bacon}, {Conseil}, {Mary}, {Brinchmann}, {Shepherd}, {Akhlaghi}, {Weilbacher}, {Piqueras}, {Wisotzki}, {Lagattuta}, {Epinat}, {Guerou}, {Inami}, {Cantalupo}, {Courbot}, {Contini}, {Richard}, {Maseda}, {Bouwens}, {Bouch{\'e}}, {Kollatschny}, {Schaye}, {Marino}, {Pello}, {Herenz}, {Guiderdoni}, \& {Carollo}}]{bacon17}
{Bacon}, R., {Conseil}, S., {Mary}, D., {et~al.} 2017, \aap, 608, A1

\bibitem[{{Bacon} {et~al.}(2021){Bacon}, {Mary}, {Garel}, {Blaizot}, {Maseda}, {Schaye}, {Wisotzki}, {Conseil}, {Brinchmann}, {Leclercq}, {Abril-Melgarejo}, {Boogaard}, {Bouch{\'e}}, {Contini}, {Feltre}, {Guiderdoni}, {Herenz}, {Kollatschny}, {Kusakabe}, {Matthee}, {Michel-Dansac}, {Nanayakkara}, {Richard}, {Roth}, {Schmidt}, {Steinmetz}, {Tresse}, {Urrutia}, {Verhamme}, {Weilbacher}, {Zabl}, \& {Zoutendijk}}]{bacon21}
{Bacon}, R., {Mary}, D., {Garel}, T., {et~al.} 2021, \aap, 647, A107

\bibitem[{{Blaizot} {et~al.}(2023){Blaizot}, {Garel}, {Verhamme}, {Katz}, {Kimm}, {Michel-Dansac}, {Mitchell}, {Rosdahl}, \& {Trebitsch}}]{blaizot23}
{Blaizot}, J., {Garel}, T., {Verhamme}, A., {et~al.} 2023, \mnras, 523, 3749

\bibitem[{{Borisova} {et~al.}(2016){Borisova}, {Cantalupo}, {Lilly}, {Marino}, {Gallego}, {Bacon}, {Blaizot}, {Bouch{\'e}}, {Brinchmann}, {Carollo}, {Caruana}, {Finley}, {Herenz}, {Richard}, {Schaye}, {Straka}, {Turner}, {Urrutia}, {Verhamme}, \& {Wisotzki}}]{borisova16}
{Borisova}, E., {Cantalupo}, S., {Lilly}, S.~J., {et~al.} 2016, \apj, 831, 39

\bibitem[{{Byrohl} {et~al.}(2021){Byrohl}, {Nelson}, {Behrens}, {Kostyuk}, {Glatzle}, {Pillepich}, {Hernquist}, {Marinacci}, \& {Vogelsberger}}]{byrohl21}
{Byrohl}, C., {Nelson}, D., {Behrens}, C., {et~al.} 2021, \mnras, 506, 5129

\bibitem[{{Cantalupo} {et~al.}(2005){Cantalupo}, {Porciani}, {Lilly}, \& {Miniati}}]{cantalupo05}
{Cantalupo}, S., {Porciani}, C., {Lilly}, S.~J., \& {Miniati}, F. 2005, \apj, 628, 61

\bibitem[{{Chang} {et~al.}(2023){Chang}, {Yang}, {Seon}, {Zabludoff}, \& {Lee}}]{chang23}
{Chang}, S.-J., {Yang}, Y., {Seon}, K.-I., {Zabludoff}, A., \& {Lee}, H.-W. 2023, \apj, 945, 100

\bibitem[{{Chen} {et~al.}(2020){Chen}, {Steidel}, {Hummels}, {Rudie}, {Dong}, {Trainor}, {Bogosavljevi{\'c}}, {Erb}, {Pettini}, {Reddy}, {Shapley}, {Strom}, {Theios}, {Faucher-Gigu{\`e}re}, {Hopkins}, \& {Kere{\v{s}}}}]{chen20}
{Chen}, Y., {Steidel}, C.~C., {Hummels}, C.~B., {et~al.} 2020, \mnras, 499, 1721

\bibitem[{{Chonis} {et~al.}(2013){Chonis}, {Blanc}, {Hill}, {Adams}, {Finkelstein}, {Gebhardt}, {Kollmeier}, {Ciardullo}, {Drory}, {Gronwall}, {Hagen}, {Overzier}, {Song}, \& {Zeimann}}]{chonis13}
{Chonis}, T.~S., {Blanc}, G.~A., {Hill}, G.~J., {et~al.} 2013, \apj, 775, 99

\bibitem[{{Chung} {et~al.}(2019){Chung}, {Dijkstra}, {Ciardi}, {Kakiichi}, \& {Naab}}]{chung19}
{Chung}, A.~S., {Dijkstra}, M., {Ciardi}, B., {Kakiichi}, K., \& {Naab}, T. 2019, \mnras, 484, 2420

\bibitem[{{Claeyssens} {et~al.}(2019){Claeyssens}, {Richard}, {Blaizot}, {Garel}, {Leclercq}, {Patr{\'\i}cio}, {Verhamme}, {Wisotzki}, {Bacon}, {Carton}, {Cl{\'e}ment}, {Herenz}, {Marino}, {Muzahid}, {Saust}, \& {Schaye}}]{claeyssens19}
{Claeyssens}, A., {Richard}, J., {Blaizot}, J., {et~al.} 2019, \mnras, 489, 5022

\bibitem[{{Dijkstra}(2014)}]{dijkstra14b}
{Dijkstra}, M. 2014, \pasa, 31, e040

\bibitem[{{Dijkstra} {et~al.}(2006){Dijkstra}, {Haiman}, \& {Spaans}}]{dijkstra06a}
{Dijkstra}, M., {Haiman}, Z., \& {Spaans}, M. 2006, \apj, 649, 14

\bibitem[{{Dijkstra} \& {Kramer}(2012)}]{dijkstra12}
{Dijkstra}, M. \& {Kramer}, R. 2012, \mnras, 424, 1672

\bibitem[{{Erb} {et~al.}(2023){Erb}, {Li}, {Steidel}, {Chen}, {Gronke}, {Strom}, {Trainor}, \& {Rudie}}]{erb22}
{Erb}, D.~K., {Li}, Z., {Steidel}, C.~C., {et~al.} 2023, \apj, 953, 118

\bibitem[{{Erb} {et~al.}(2018){Erb}, {Steidel}, \& {Chen}}]{erb18}
{Erb}, D.~K., {Steidel}, C.~C., \& {Chen}, Y. 2018, \apjl, 862, L10

\bibitem[{{Erb} {et~al.}(2014){Erb}, {Steidel}, {Trainor}, {Bogosavljevi{\'c}}, {Shapley}, {Nestor}, {Kulas}, {Law}, {Strom}, {Rudie}, {Reddy}, {Pettini}, {Konidaris}, {Mace}, {Matthews}, \& {McLean}}]{erb14}
{Erb}, D.~K., {Steidel}, C.~C., {Trainor}, R.~F., {et~al.} 2014, \apj, 795, 33

\bibitem[{{Gallego} {et~al.}(2021){Gallego}, {Cantalupo}, {Sarpas}, {Duboeuf}, {Lilly}, {Pezzulli}, {Marino}, {Matthee}, {Wisotzki}, {Schaye}, {Richard}, {Kusakabe}, \& {Mauerhofer}}]{gallego21}
{Gallego}, S.~G., {Cantalupo}, S., {Sarpas}, S., {et~al.} 2021, \mnras, 504, 16

\bibitem[{{Garel} {et~al.}(2015){Garel}, {Blaizot}, {Guiderdoni}, {Michel-Dansac}, {Hayes}, \& {Verhamme}}]{garel15}
{Garel}, T., {Blaizot}, J., {Guiderdoni}, B., {et~al.} 2015, \mnras, 450, 1279

\bibitem[{{Girelli} {et~al.}(2020){Girelli}, {Pozzetti}, {Bolzonella}, {Giocoli}, {Marulli}, \& {Baldi}}]{girelli20}
{Girelli}, G., {Pozzetti}, L., {Bolzonella}, M., {et~al.} 2020, \aap, 634, A135

\bibitem[{{Gould} \& {Weinberg}(1996)}]{gould96}
{Gould}, A. \& {Weinberg}, D.~H. 1996, \apj, 468, 462

\bibitem[{{Gronke}(2017)}]{gronke17}
{Gronke}, M. 2017, \aap, 608, A139

\bibitem[{{Gronwall} {et~al.}(2007){Gronwall}, {Ciardullo}, {Hickey}, {Gawiser}, {Feldmeier}, {van Dokkum}, {Urry}, {Herrera}, {Lehmer}, {Infante}, {Orsi}, {Marchesini}, {Blanc}, {Francke}, {Lira}, \& {Treister}}]{gronwall07}
{Gronwall}, C., {Ciardullo}, R., {Hickey}, T., {et~al.} 2007, \apj, 667, 79

\bibitem[{Guo {et~al.}(2023)Guo, Bacon, Wisotzki, Garel, Blaizot, Schaye, Richard, Alonso, Leclercq, Boogaard, Kusakabe, \& Pharo}]{guo23}
Guo, Y., Bacon, R., Wisotzki, L., {et~al.} 2023, arXiv:2309.05513

\bibitem[{{Guo} {et~al.}(2020{\natexlab{a}}){Guo}, {Jiang}, {Egami}, {Ning}, {Zheng}, \& {Ho}}]{guo20b}
{Guo}, Y., {Jiang}, L., {Egami}, E., {et~al.} 2020{\natexlab{a}}, \apj, 902, 137

\bibitem[{{Guo} {et~al.}(2020{\natexlab{b}}){Guo}, {Maiolino}, {Jiang}, {Matsuoka}, {Nagao}, {Dors}, {Ginolfi}, {Henden}, {Bennett}, {Sijacki}, \& {Puchwein}}]{guo20}
{Guo}, Y., {Maiolino}, R., {Jiang}, L., {et~al.} 2020{\natexlab{b}}, \apj, 898, 26

\bibitem[{{Haiman} {et~al.}(2000){Haiman}, {Spaans}, \& {Quataert}}]{haiman00}
{Haiman}, Z., {Spaans}, M., \& {Quataert}, E. 2000, \apjl, 537, L5

\bibitem[{{Hashimoto} {et~al.}(2015){Hashimoto}, {Verhamme}, {Ouchi}, {Shimasaku}, {Schaerer}, {Nakajima}, {Shibuya}, {Rauch}, {Ono}, \& {Goto}}]{hashimoto15}
{Hashimoto}, T., {Verhamme}, A., {Ouchi}, M., {et~al.} 2015, \apj, 812, 157

\bibitem[{{Hayashino} {et~al.}(2004){Hayashino}, {Matsuda}, {Tamura}, {Yamauchi}, {Yamada}, {Ajiki}, {Fujita}, {Murayama}, {Nagao}, {Ohta}, {Okamura}, {Ouchi}, {Shimasaku}, {Shioya}, \& {Taniguchi}}]{hayashino04}
{Hayashino}, T., {Matsuda}, Y., {Tamura}, H., {et~al.} 2004, \aj, 128, 2073

\bibitem[{{Hayes} {et~al.}(2021){Hayes}, {Runnholm}, {Gronke}, \& {Scarlata}}]{hayes21}
{Hayes}, M.~J., {Runnholm}, A., {Gronke}, M., \& {Scarlata}, C. 2021, \apj, 908, 36

\bibitem[{{Hopkins} {et~al.}(2018){Hopkins}, {Wetzel}, {Kere{\v{s}}}, {Faucher-Gigu{\`e}re}, {Quataert}, {Boylan-Kolchin}, {Murray}, {Hayward}, {Garrison-Kimmel}, {Hummels}, {Feldmann}, {Torrey}, {Ma}, {Angl{\'e}s-Alc{\'a}zar}, {Su}, {Orr}, {Schmitz}, {Escala}, {Sanderson}, {Grudi{\'c}}, {Hafen}, {Kim}, {Fitts}, {Bullock}, {Wheeler}, {Chan}, {Elbert}, \& {Narayanan}}]{hopkins18}
{Hopkins}, P.~F., {Wetzel}, A., {Kere{\v{s}}}, D., {et~al.} 2018, \mnras, 480, 800

\bibitem[{{Inoue} {et~al.}(2014){Inoue}, {Shimizu}, {Iwata}, \& {Tanaka}}]{inoue14}
{Inoue}, A.~K., {Shimizu}, I., {Iwata}, I., \& {Tanaka}, M. 2014, \mnras, 442, 1805

\bibitem[{{Kikuchihara} {et~al.}(2022){Kikuchihara}, {Harikane}, {Ouchi}, {Ono}, {Shibuya}, {Itoh}, {Kakuma}, {Inoue}, {Kusakabe}, {Shimasaku}, {Momose}, {Sugahara}, {Kikuta}, {Saito}, {Kashikawa}, {Zhang}, \& {Lee}}]{kikuchi22}
{Kikuchihara}, S., {Harikane}, Y., {Ouchi}, M., {et~al.} 2022, \apj, 931, 97

\bibitem[{{Kikuta} {et~al.}(2023){Kikuta}, {Ouchi}, {Shibuya}, {Liang}, {Umeda}, {Matsumoto}, {Shimasaku}, {Harikane}, {Ono}, {Inoue}, {Yamanaka}, {Kusakabe}, {Momose}, {Kashikawa}, {Matsuda}, \& {Lee}}]{kikuta23}
{Kikuta}, S., {Ouchi}, M., {Shibuya}, T., {et~al.} 2023, \apjs, 268, 24

\bibitem[{{Kusakabe} {et~al.}(2022){Kusakabe}, {Verhamme}, {Blaizot}, {Garel}, {Wisotzki}, {Leclercq}, {Bacon}, {Schaye}, {Gallego}, {Kerutt}, {Matthee}, {Maseda}, {Nanayakkara}, {Pell{\'o}}, {Richard}, {Tresse}, {Urrutia}, \& {Vitte}}]{kusakabe22}
{Kusakabe}, H., {Verhamme}, A., {Blaizot}, J., {et~al.} 2022, \aap, 660, A44

\bibitem[{{Laursen} {et~al.}(2011){Laursen}, {Sommer-Larsen}, \& {Razoumov}}]{laursen11}
{Laursen}, P., {Sommer-Larsen}, J., \& {Razoumov}, A.~O. 2011, \apj, 728, 52

\bibitem[{{Leclercq} {et~al.}(2020){Leclercq}, {Bacon}, {Verhamme}, {Garel}, {Blaizot}, {Brinchmann}, {Cantalupo}, {Claeyssens}, {Conseil}, {Contini}, {Hashimoto}, {Herenz}, {Kusakabe}, {Marino}, {Maseda}, {Matthee}, {Mitchell}, {Pezzulli}, {Richard}, {Schmidt}, \& {Wisotzki}}]{leclercq20}
{Leclercq}, F., {Bacon}, R., {Verhamme}, A., {et~al.} 2020, \aap, 635, A82

\bibitem[{{Leclercq} {et~al.}(2017){Leclercq}, {Bacon}, {Wisotzki}, {Mitchell}, {Garel}, {Verhamme}, {Blaizot}, {Hashimoto}, {Herenz}, {Conseil}, {Cantalupo}, {Inami}, {Contini}, {Richard}, {Maseda}, {Schaye}, {Marino}, {Akhlaghi}, {Brinchmann}, \& {Carollo}}]{leclercq17}
{Leclercq}, F., {Bacon}, R., {Wisotzki}, L., {et~al.} 2017, \aap, 608, A8

\bibitem[{{Li} {et~al.}(2022){Li}, {Steidel}, {Gronke}, {Chen}, \& {Matsuda}}]{li22}
{Li}, Z., {Steidel}, C.~C., {Gronke}, M., {Chen}, Y., \& {Matsuda}, Y. 2022, \mnras, 513, 3414

\bibitem[{{Lujan Niemeyer} {et~al.}(2022){Lujan Niemeyer}, {Komatsu}, {Byrohl}, {Davis}, {Fabricius}, {Gebhardt}, {Hill}, {Wisotzki}, {Bowman}, {Ciardullo}, {Farrow}, {Finkelstein}, {Gawiser}, {Gronwall}, {Jeong}, {Landriau}, {Liu}, {Cooper}, {Ouchi}, {Schneider}, \& {Zeimann}}]{maja22a}
{Lujan Niemeyer}, M., {Komatsu}, E., {Byrohl}, C., {et~al.} 2022, \apj, 929, 90

\bibitem[{{Martin} {et~al.}(2010){Martin}, {Moore}, {Morrissey}, {Matuszewski}, {Rahman}, {Adkins}, \& {Epps}}]{martin10}
{Martin}, C., {Moore}, A., {Morrissey}, P., {et~al.} 2010, in Society of Photo-Optical Instrumentation Engineers (SPIE) Conference Series, Vol. 7735, Ground-based and Airborne Instrumentation for Astronomy III, ed. I.~S. {McLean}, S.~K. {Ramsay}, \& H.~{Takami}, 77350M

\bibitem[{{Mas-Ribas} \& {Dijkstra}(2016)}]{ribas16}
{Mas-Ribas}, L. \& {Dijkstra}, M. 2016, \apj, 822, 84

\bibitem[{{Mas-Ribas} {et~al.}(2017){Mas-Ribas}, {Dijkstra}, {Hennawi}, {Trenti}, {Momose}, \& {Ouchi}}]{mas17}
{Mas-Ribas}, L., {Dijkstra}, M., {Hennawi}, J.~F., {et~al.} 2017, \apj, 841, 19

\bibitem[{{McLinden} {et~al.}(2011){McLinden}, {Finkelstein}, {Rhoads}, {Malhotra}, {Hibon}, {Richardson}, {Cresci}, {Quirrenbach}, {Pasquali}, {Bian}, {Fan}, \& {Woodward}}]{mclinden11}
{McLinden}, E.~M., {Finkelstein}, S.~L., {Rhoads}, J.~E., {et~al.} 2011, \apj, 730, 136

\bibitem[{{Mitchell} {et~al.}(2021){Mitchell}, {Blaizot}, {Cadiou}, {Dubois}, {Garel}, \& {Rosdahl}}]{mitchell20}
{Mitchell}, P.~D., {Blaizot}, J., {Cadiou}, C., {et~al.} 2021, \mnras, 501, 5757

\bibitem[{{Momose} {et~al.}(2014){Momose}, {Ouchi}, {Nakajima}, {Ono}, {Shibuya}, {Shimasaku}, {Yuma}, {Mori}, \& {Umemura}}]{momose14}
{Momose}, R., {Ouchi}, M., {Nakajima}, K., {et~al.} 2014, \mnras, 442, 110

\bibitem[{{Momose} {et~al.}(2016){Momose}, {Ouchi}, {Nakajima}, {Ono}, {Shibuya}, {Shimasaku}, {Yuma}, {Mori}, \& {Umemura}}]{momose16}
{Momose}, R., {Ouchi}, M., {Nakajima}, K., {et~al.} 2016, \mnras, 457, 2318

\bibitem[{{Morrissey} {et~al.}(2012){Morrissey}, {Matuszewski}, {Martin}, {Moore}, {Adkins}, {Epps}, {Bartos}, {Cabak}, {Cowley}, {Davis}, {Delacroix}, {Fucik}, {Hilliard}, {James}, {Kaye}, {Lingner}, {Neill}, {Pistor}, {Phillips}, {Rockosi}, \& {Weber}}]{morrissey12}
{Morrissey}, P., {Matuszewski}, M., {Martin}, C., {et~al.} 2012, in Society of Photo-Optical Instrumentation Engineers (SPIE) Conference Series, Vol. 8446, Ground-based and Airborne Instrumentation for Astronomy IV, ed. I.~S. {McLean}, S.~K. {Ramsay}, \& H.~{Takami}, 844613

\bibitem[{{Muzahid} {et~al.}(2020){Muzahid}, {Schaye}, {Marino}, {Cantalupo}, {Brinchmann}, {Contini}, {Wendt}, {Wisotzki}, {Zabl}, {Bouch{\'e}}, {Akhlaghi}, {Chen}, {Claeyssens}, {Johnson}, {Leclercq}, {Maseda}, {Matthee}, {Richard}, {Urrutia}, \& {Verhamme}}]{muzahid20}
{Muzahid}, S., {Schaye}, J., {Marino}, R.~A., {et~al.} 2020, \mnras, 496, 1013

\bibitem[{{Ouchi} {et~al.}(2020){Ouchi}, {Ono}, \& {Shibuya}}]{ouchi20}
{Ouchi}, M., {Ono}, Y., \& {Shibuya}, T. 2020, \araa, 58, 617

\bibitem[{{Ouchi} {et~al.}(2008){Ouchi}, {Shimasaku}, {Akiyama}, {Simpson}, {Saito}, {Ueda}, {Furusawa}, {Sekiguchi}, {Yamada}, {Kodama}, {Kashikawa}, {Okamura}, {Iye}, {Takata}, {Yoshida}, \& {Yoshida}}]{ouchi08}
{Ouchi}, M., {Shimasaku}, K., {Akiyama}, M., {et~al.} 2008, \apjs, 176, 301

\bibitem[{{Patr{\'\i}cio} {et~al.}(2016){Patr{\'\i}cio}, {Richard}, {Verhamme}, {Wisotzki}, {Brinchmann}, {Turner}, {Christensen}, {Weilbacher}, {Blaizot}, {Bacon}, {Contini}, {Lagattuta}, {Cantalupo}, {Cl{\'e}ment}, \& {Soucail}}]{Patricio16}
{Patr{\'\i}cio}, V., {Richard}, J., {Verhamme}, A., {et~al.} 2016, \mnras, 456, 4191

\bibitem[{{Rauch} {et~al.}(2008){Rauch}, {Haehnelt}, {Bunker}, {Becker}, {Marleau}, {Graham}, {Cristiani}, {Jarvis}, {Lacey}, {Morris}, {Peroux}, {R{\"o}ttgering}, \& {Theuns}}]{rauch08}
{Rauch}, M., {Haehnelt}, M., {Bunker}, A., {et~al.} 2008, \apj, 681, 856

\bibitem[{{Richard} {et~al.}(2021){Richard}, {Claeyssens}, {Lagattuta}, {Guaita}, {Bauer}, {Pello}, {Carton}, {Bacon}, {Soucail}, {Lyon}, {Kneib}, {Mahler}, {Cl{\'e}ment}, {Mercier}, {Variu}, {Tamone}, {Ebeling}, {Schmidt}, {Nanayakkara}, {Maseda}, {Weilbacher}, {Bouch{\'e}}, {Bouwens}, {Wisotzki}, {de la Vieuville}, {Martinez}, \& {Patr{\'\i}cio}}]{richard21}
{Richard}, J., {Claeyssens}, A., {Lagattuta}, D., {et~al.} 2021, \aap, 646, A83

\bibitem[{{Rosdahl} \& {Blaizot}(2012)}]{rosdahl12}
{Rosdahl}, J. \& {Blaizot}, J. 2012, \mnras, 423, 344

\bibitem[{{Shibuya} {et~al.}(2012){Shibuya}, {Kashikawa}, {Ota}, {Iye}, {Ouchi}, {Furusawa}, {Shimasaku}, \& {Hattori}}]{shibuya12}
{Shibuya}, T., {Kashikawa}, N., {Ota}, K., {et~al.} 2012, \apj, 752, 114

\bibitem[{{Shibuya} {et~al.}(2014){Shibuya}, {Ouchi}, {Nakajima}, {Hashimoto}, {Ono}, {Rauch}, {Gauthier}, {Shimasaku}, {Goto}, {Mori}, \& {Umemura.}}]{shibuya14}
{Shibuya}, T., {Ouchi}, M., {Nakajima}, K., {et~al.} 2014, \apj, 788, 74

\bibitem[{{Solimano} {et~al.}(2022){Solimano}, {Gonz{\'a}lez-L{\'o}pez}, {Aravena}, {Johnston}, {Moya-Sierralta}, {Barrientos}, {Bayliss}, {Gladders}, {Infante}, {Ledoux}, {L{\'o}pez}, {Poudel}, {Rigby}, {Sharon}, \& {Tejos}}]{solimano22}
{Solimano}, M., {Gonz{\'a}lez-L{\'o}pez}, J., {Aravena}, M., {et~al.} 2022, \apj, 935, 17

\bibitem[{{Song} {et~al.}(2020){Song}, {Seon}, \& {Hwang}}]{song20}
{Song}, H., {Seon}, K.-I., \& {Hwang}, H.~S. 2020, \apj, 901, 41

\bibitem[{{Steidel} {et~al.}(2011){Steidel}, {Bogosavljevi{\'c}}, {Shapley}, {Kollmeier}, {Reddy}, {Erb}, \& {Pettini}}]{steidel11}
{Steidel}, C.~C., {Bogosavljevi{\'c}}, M., {Shapley}, A.~E., {et~al.} 2011, \apj, 736, 160

\bibitem[{{Thompson} \& {Heckman}(2024)}]{thompson24}
{Thompson}, T.~A. \& {Heckman}, T.~M. 2024, arXiv e-prints, arXiv:2406.08561

\bibitem[{{Tumlinson} {et~al.}(2017){Tumlinson}, {Peeples}, \& {Werk}}]{tumlinson17}
{Tumlinson}, J., {Peeples}, M.~S., \& {Werk}, J.~K. 2017, \araa, 55, 389

\bibitem[{{Verhamme} {et~al.}(2018){Verhamme}, {Garel}, {Ventou}, {Contini}, {Bouch{\'e}}, {Herenz}, {Richard}, {Bacon}, {Schmidt}, {Maseda}, {Marino}, {Brinchmann}, {Cantalupo}, {Caruana}, {Cl{\'e}ment}, {Diener}, {Drake}, {Hashimoto}, {Inami}, {Kerutt}, {Kollatschny}, {Leclercq}, {Patr{\'\i}cio}, {Schaye}, {Wisotzki}, \& {Zabl}}]{verhamme18}
{Verhamme}, A., {Garel}, T., {Ventou}, E., {et~al.} 2018, \mnras, 478, L60

\bibitem[{{Verhamme} {et~al.}(2006){Verhamme}, {Schaerer}, \& {Maselli}}]{verhamme06}
{Verhamme}, A., {Schaerer}, D., \& {Maselli}, A. 2006, \aap, 460, 397

\bibitem[{{Wisotzki} {et~al.}(2016){Wisotzki}, {Bacon}, {Blaizot}, {Brinchmann}, {Herenz}, {Schaye}, {Bouch{\'e}}, {Cantalupo}, {Contini}, {Carollo}, {Caruana}, {Courbot}, {Emsellem}, {Kamann}, {Kerutt}, {Leclercq}, {Lilly}, {Patr{\'\i}cio}, {Sandin}, {Steinmetz}, {Straka}, {Urrutia}, {Verhamme}, {Weilbacher}, \& {Wendt}}]{wisotzki16}
{Wisotzki}, L., {Bacon}, R., {Blaizot}, J., {et~al.} 2016, \aap, 587, A98

\bibitem[{{Wisotzki} {et~al.}(2018){Wisotzki}, {Bacon}, {Brinchmann}, {Cantalupo}, {Richter}, {Schaye}, {Schmidt}, {Urrutia}, {Weilbacher}, {Akhlaghi}, {Bouch{\'e}}, {Contini}, {Guiderdoni}, {Herenz}, {Inami}, {Kerutt}, {Leclercq}, {Marino}, {Maseda}, {Monreal-Ibero}, {Nanayakkara}, {Richard}, {Saust}, {Steinmetz}, \& {Wendt}}]{wisotzki18}
{Wisotzki}, L., {Bacon}, R., {Brinchmann}, J., {et~al.} 2018, \nat, 562, 229

\bibitem[{{Yamada} {et~al.}(2012){Yamada}, {Matsuda}, {Kousai}, {Hayashino}, {Morimoto}, \& {Umemura}}]{yamada12}
{Yamada}, T., {Matsuda}, Y., {Kousai}, K., {et~al.} 2012, \apj, 751, 29

\bibitem[{{Zheng} \& {Miralda-Escud{\'e}}(2002)}]{zheng02}
{Zheng}, Z. \& {Miralda-Escud{\'e}}, J. 2002, \apj, 578, 33

\end{thebibliography}

%%%%%%%%%%%%%%%%%%%%%%%%%%%%%%%%%%%%%%%%%%%%%%%%%%%%%%%%%%%%%%%%%%%%
\begin{appendix} %First  appendix

\section{The \lya\ surface brightness profile}
\label{appendix:sb_profile}

\begin{figure*}[]
\centering
\includegraphics[width=0.45\textwidth]{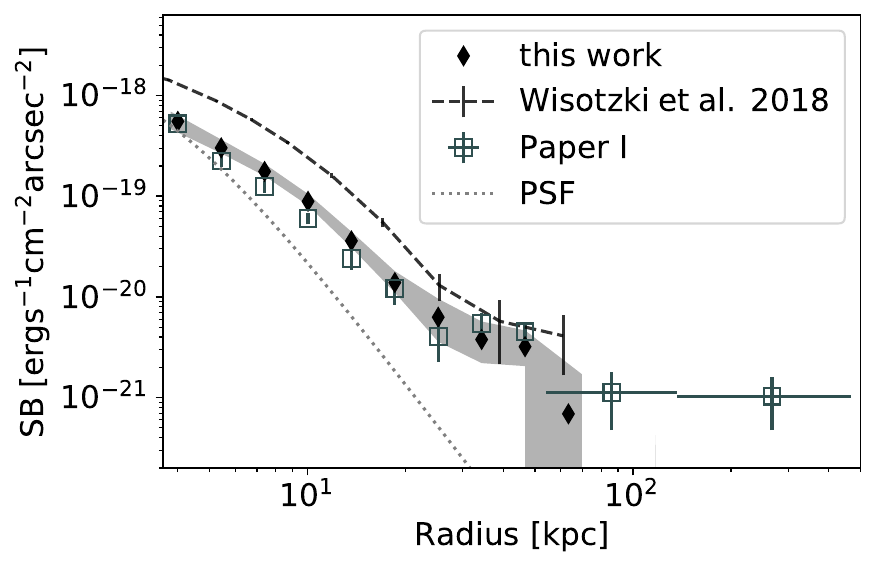}
\includegraphics[width=0.45\textwidth]{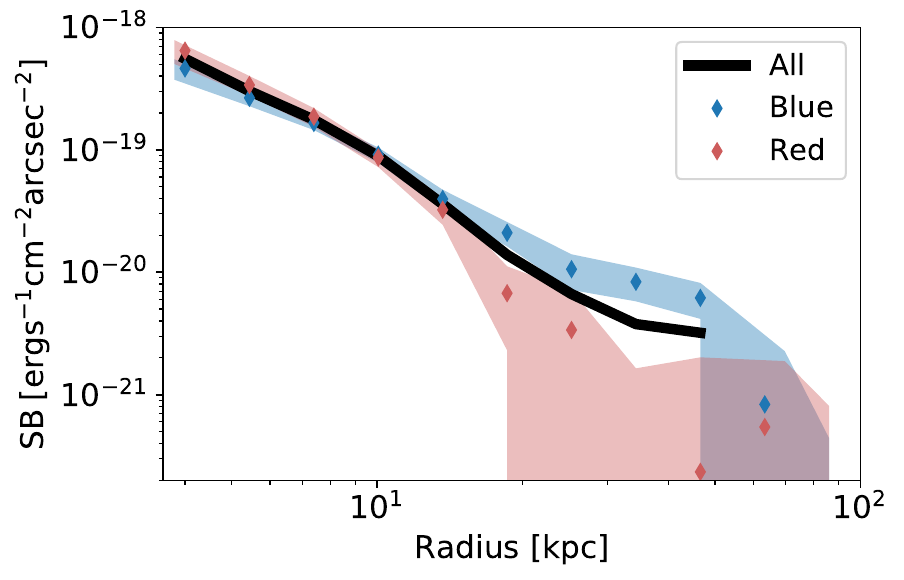}
\caption{ The \lya\ surface brightness profile.
\textit{Left:} The black symbols show the \lya\ surface brightness profile extracted within a pseudo-NB of [-600, 600]~km/s, with the centroid of the pseudo-NB at the \lya\ peak of the target LAE. The grey shadow indicates the error range. The open squares show the \lya\ surface brightness profile in Paper~I. The dashed line shows the \lya\ surface brightness profile in \citet{wisotzki18}. The grey dashed line shows the PSF profile.
\textit{Right:} The black curve corresponds to the black symbols in the left panel. The red and blue dots with error ranges present the \lya\ surface brightness profile extracted by pseudo-NB [0, 600]~km/s and [-600, 0]~km/s, respectively. These two profiles are multiplied by a factor of 2 for better visualisation.
\label{fig_Lya_sb_profile}}
\end{figure*}

In Figure~\ref{fig_Lya_sb_profile}, we present the \lya\ surface brightness profile measured from the median-stacked datacube.
We use a pseudo-NB of [-600, 600]~km/s with the centroid at the central \lya\ line peak.
In the left panel of Figure~\ref{fig_Lya_sb_profile}, we also compare this work with the previous analysis and the PSF of MUSE.
The \lya\ surface brightness profile agrees well with Paper~I, which is based on the same data sample but adopt a different stacking strategy:
Paper~I stack the \lya\ surface brightness profiles of individual galaxies extracted from pseudo-NB images, while this work stacks the datacube directly.
The \lya\ surface brightness profile has similar shape to that of \citet{wisotzki18} but with different scaling. \citet{wisotzki18} is based on a brighter LAE sample.
The interpretation of the shape of the \lya\ surface brightness profile is beyond the scope of this paper.
We refer the readers to Paper~I and \citet{wisotzki18} for a detailed analysis and comparison with other literature results.

In the right panel of Figure~\ref{fig_Lya_sb_profile}, we present the \lya\ surface brightness profiles from the bluer and redder wavelengths relative to the central \lya\ line peak.
The profiles are extracted by pseudo-NB with width of 600~km/s, and multiplied by a factor of 2 for better visualization.
The blueshift of \lya\ line with respect to the central \lya\ line peak naturally produce different behaviour of \lya\ surface brightness profiles: the blue surface brightness profile dominates the large distances, and produces a flattening trend.

%\begin{comment}
%\end{comment}

\section{The spectral morphology of the LAHs at $4<z<5$}
\label{appendix:z45}
\begin{figure}[]
\centering
\includegraphics[width=0.45\textwidth]{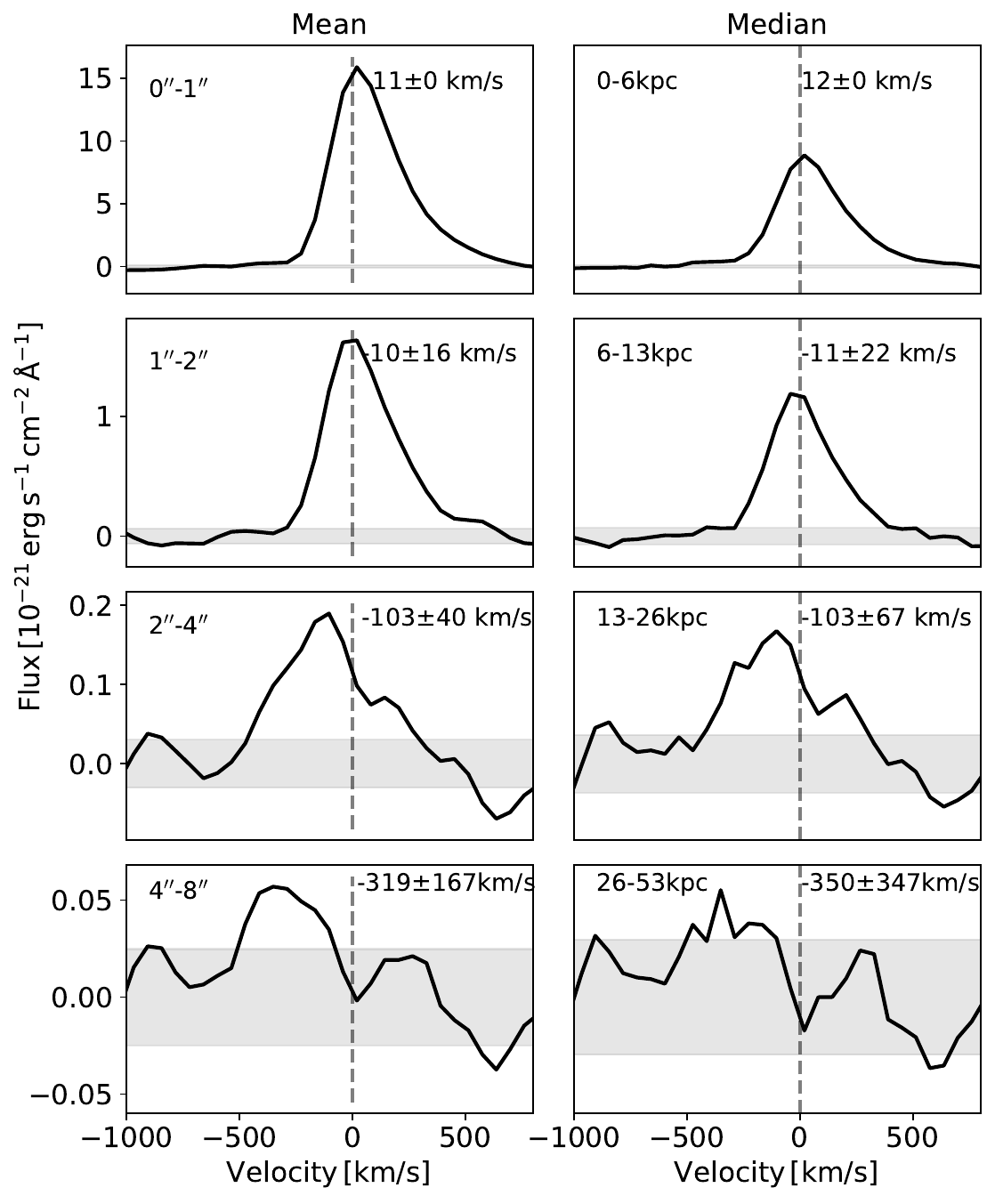}
\caption{ Same as Figure~\ref{fig_Lya_profile}, but for LAEs at $4<z<5$.
\label{fig_Lya_profile_z2}}
\end{figure}

In Figure~\ref{fig_Lya_profile_z2}, we present the stack of 128 MXDF LAEs at $4<z<5$.
The stack shows a similar large-scale pattern as in $3<z<4$ (Figure~\ref{fig_Lya_profile}), but with a lower S/N.

%\begin{comment}
%\end{comment}

\end{appendix}

%%%%%%%%%%%%%%%%%%%%%%%%%%%%%%%%%%%%%%%%%%%%%%%%%%%%%%%%%%%%%%%%%%%%
\end{document}